# Prospects for Acoustically Monitoring Ecosystem Tipping Points

## Authors:

Neel P. Le Penru [1,2]*, Thomas M. Bury [3], Sarab S. Sethi [4], Robert M. Ewers [2], Lorenzo Picinali [1]

## Affiliations:

1. Audio Experience Design, Dyson School of Design Engineering, Imperial College London, London, UK
2. Georgina Mace Centre for the Living Planet, Imperial College London, Ascot, UK
3. Department of Physiology, McGill University, Montreal, Canada
4. Department of Life Sciences, Imperial College London, London UK

**\* Corresponding author – contact information:**
*Email:* neel.le-penru16@imperial.ac.uk

# Abstract

Many ecosystems can undergo important qualitative changes, including sudden transitions to alternative stable states, in response to perturbations or increments in conditions. Such 'tipping points' are often preceded by declines in aspects of ecosystem resilience, namely the capacity to recover from perturbations, that leave various spatial and temporal signatures. These so-called 'early warning signals' have been used to anticipate transitions in diverse real systems, but many of the high-throughput, autonomous monitoring technologies that are transforming ecology have yet to be fully leveraged to this end. Acoustic monitoring in particular is a powerful tool for quantifying biodiversity, tracking ecosystem health, and facilitating conservation. By deploying acoustic recorders in diverse environments, researchers have gained insights from the calls and behaviour of individual species to higher-level soundscape features that describe habitat quality and even predict species occurrence. Here, we draw on theory and practice to advocate for using acoustics to probe ecosystem resilience and identify emerging and established early warning signals of tipping points. With a focus on pragmatic considerations, we emphasise that despite limits to tipping point theory and the current scale and transferability of data, acoustics could be instrumental in understanding resilience and tipping potential across distinct ecosystems and scales.

# Keywords

# 1. Introduction

Across diverse domains and scales, dynamical systems are known to possess multiple stable states between which they may transition (Scheffer, 2009; Strogatz, 2024). Beams buckle, populations explode or collapse, financial markets crash, human brains experience seizures or different psychological states, ice ages begin and end, and entire ecosystems small and vast can transform drastically faster than expected. Such transitions, sometimes termed 'regime shifts' (Folke et al., 2004; Carpenter et al., 2011; Dakos et al., 2015; Sánchez-Pinillos et al., 2024) or 'critical transitions' (Scheffer, 2009; Scheffer et al., 2009), can arise due to various underlying mechanisms that may produce gradual, fast, or sudden, nonlinear shifts. 'Tipping points' denote those caused by small perturbations amplified by positive feedback (Lenton et al., 2008; Dakos et al., 2024), extending the term popularised by Gladwell (2000) (though its use has varied somewhat in the past (Scheffer et al., 2009; van Nes et al., 2016)). Tipping points produce abrupt, qualitative change to the state of a system that may be difficult or impossible to reverse (Dakos et al., 2024; Lenton et al., 2008). As such, they have been widely researched in recent decades, especially in climate science (Lenton et al., 2008; Liu et al., 2017; Armstrong McKay et al., 2022; Dakos et al., 2024), ecology (Scheffer et al., 2001; Folke et al., 2004; Carpenter et al., 2011; Hirota et al., 2011; Lenton, 2013; Kéfi et al., 2014; Scheffer et al., 2015; Boulton et al., 2022; Flores et al., 2024) and their intersection in Earth system science (Rocha et al., 2018; Boers and Rypdal, 2021; Lenton et al., 2022, 2024; Westen et al., 2024), where the risk to natural systems and humanity is particularly great (Lenton et al., 2019). Despite the urgent need to monitor and anticipate tipping points, many advancing autonomous monitoring technologies that empower data-driven ecology have not yet been applied to real-time assessment of impending ecosystem tipping points. Here, we spotlight acoustic monitoring as a particularly insightful technology to which existing tipping point early warnings could be readily extended for enhanced understanding of whether and how ecosystems under intensifying environmental change worldwide may tip.

Tipping points are predominantly brought about by bifurcations (Fig. 1; Box 1), noise that is internal or due to stochastic forcing pushing the system out of its current stable state's 'basin of attraction' (Fig. 1; Box 1), or, sometimes, rapid forcing in excess of a threshold rate (Ashwin et al., 2012; Dakos et al., 2024; Lenton, 2013). Bifurcations are a sudden qualitative change to the dynamics of a system – typically, the loss of the current state's stability, resulting in a shift to an alternative stable state – in response to gradual change in the control parameter(s) of the system passing some critical threshold (a.k.a. the bifurcation point) (Dakos et al., 2024; Lenton, 2013; Strogatz, 2024). Much of the work on early warning signals for tipping points has focussed on bifurcations as they are widespread and satisfy universal mathematical laws that can be detected in temporal and spatial data (Scheffer et al., 2009).

Moreover, according to the centre manifold theorem, the dynamics of complex systems near a local bifurcation simplify to a subset of lower-dimensional, prototypical 'normal forms' (Kuznetsov, 2023; Strogatz, 2024), promoting a strong basis for detecting generic EWS in real systems. These EWS generally exploit the declining stability before a bifurcation, which causes critical slowing down of a system's response to perturbation (Scheffer et al., 2009; Strogatz, 2024; Wissel, 1984), yielding increases in variance and autocorrelation – two of the most widespread EWS. The capacity to recover from perturbations is often also considered integral to a system's resilience – though this has long been discussed in ecology (Holling, 1973, 1996; Carpenter et al., 2001; Hodgson et al., 2015; Walker, 2020; Dakos and Kéfi, 2022). In addition, noise-induced tipping and bifurcations often work together in reality, since the weakening of a current state's stability (negative feedback; Box 1) towards a bifurcation increases the chances that noise or a typical perturbation can push the system to an alternate attractor – sometimes with devastating consequences. EWS can thus indicate increased risk of tipping as a system loses (aspects of) resilience, spurring prolific research into their development and application (Dakos et al., 2024).

In ecology, the notion that dynamical natural systems from populations to whole ecosystems may possess multiple stable states has long existed (Lotka, 1925). The concept of nonlinear state transitions and their thresholds gained traction from the 1960s (Holling, 1973; Lewontin, 1969; May, 1977) after they were described in simple ecological models, namely of grazing (Noy-Meir, 1975) and harvesting systems (Lotka, 1925; May, 1977), predator-prey interactions (Rosenzweig and MacArthur, 1963), and insect outbreaks (May, 1977). The observation of large-scale, abrupt regime shifts in various real ecosystems – including clear lakes becoming turbid, macroalgae dominating coral reefs, forests switching to grasslands, and arid environment desertification – also paved the way for further research on characterising and anticipating such transitions (Scheffer et al., 2001). Since early work suggested critical slowing down could be measured in fluctuations of the system state caused by stochastic forcing (Ives, 1995; Kubo, 1966) and models revealed that critical slowing down increases autocorrelation (Held and Kleinen, 2004; Ives, 1995) and variance (Carpenter and Brock, 2006), these metrics and other EWS have been identified in a fast-growing number of systems (see reviews in (Clements and Ozgul, 2018; Dakos et al., 2024; Scheffer et al., 2015)).

A key challenge in detecting EWS in real ecosystems is the limited data of sufficient spatial and temporal scale, quality, and resolution (Clements et al., 2015; Clements and Ozgul, 2018). While EWS of past climatic shifts have been demonstrated in paleoclimate data (Dakos et al., 2008; Livina et al., 2010), historical data for ecosystems is less forthcoming. One solution is to study various spatial patterns, such as spatial variance and autocorrelation (Guttal and Jayaprakash, 2009), or vegetation patch size and distribution (relevant to e.g., self-organisation of semi-arid vegetation under a drying climate) (Kéfi et al., 2014). Another is space-for-time substitution: monitoring ecosystems along spatial gradients of human, environmental, and/or climatic pressures (Ciemer et al., 2019; de Oliveira Roque et al., 2018; Eby et al., 2017; Hirota et al., 2011). In the past 15 years, satellite remote sensing has also emerged as a vital tool for ecosystem monitoring as it provides exceptional resolution and coverage of the Earth, and certain sensors now have 50-yr records (Lenton et al., 2024). Remote sensing has thus become the predominant data source for ecosystem tipping point research (Dakos et al., 2024) and has been leveraged to identify temporal and/or spatial EWS in various ecosystems globally, including the Amazon (Boulton et al., 2022), other tropical rainforests (Hirota et al., 2011; Verbesselt et al., 2016) and even the biosphere at large (Lenton et al., 2022).

The availability of ecological field survey data has, however, grown significantly in recent years, thanks to the emergence of various high-throughput, autonomous monitoring technologies (Besson et al., 2022), shaping the view that ecology is increasingly a "big data science" (Farley et al., 2018). These technologies include acoustic sensors, environmental DNA sensing, camera traps and other image/electromagnetic-based systems such as thermal, infrared and LiDAR sensors (Besson et al., 2022; Pimm et al., 2015). Numerous studies have leveraged data from these sensing modalities, but we argue there remain considerable untapped opportunities to search within large and expanding monitoring datasets for EWS and integrate them, alongside satellite sensors, into multimodal, holistic models of ecosystems for more comprehensive assessment of ecosystem health, dynamics, resilience, and proximity to tipping.

We contend that ecoacoustics – the analysis of natural soundscapes – presents a particularly promising avenue for monitoring ecosystem tipping points, for several overarching reasons. First, it is information rich. As sound propagates from an emitter in all directions, past many barriers and over fairly large areas, acoustic monitoring can capture information outside the field of view or spatial reach of other sensors. Second, it is relatively inexpensive, particularly in terms of the declining cost of recording hardware. Third, it provides a high temporal resolution record for examining ecosystems at many levels, from individuals to populations and higher-level metrics of overall ecosystem health (Ross et al., 2023; Sueur and Farina, 2015). By deploying acoustic autonomous recording units, researchers have undertaken the passive acoustic monitoring of diverse ecosystems, uncovering a wealth of ecological, behavioural, and conservation-relevant insights across the globe (Darras et al., 2025) on larger scales and with less bias and disturbance than manual field surveys (Gibb et al., 2019; Sugai et al., 2019; Wrege et al., 2017). Passive acoustic monitoring has been used to study fish (Van Parijs et al., 2009), cetaceans (Rice et al., 2021), frogs (Measey et al., 2017), insects (Kohlberg et al., 2024), birds (Symes et al., 2022), terrestrial mammals (Clink and Klinck, 2021; Wrege et al., 2017) and facilitate estimates of overall biodiversity and habitat quality (Sethi et al., 2020b). Larger-scale passive acoustic monitoring deployments are also increasing, capturing recordings over landscapes (Ross et al., 2018; Wood et al., 2019), countries (Sethi et al., 2021) and continents (Roe et al., 2021). Further, passive acoustic monitoring has captured aspects of stability and resilience by revealing various ecosystem responses to disturbances including human infrastructure development (Deichmann et al., 2017), forest fragmentation (Burivalova et al., 2018), species invasion (Gasc et al., 2018), fires (Wood et al., 2024), and typhoons/hurricanes (Gottesman et al., 2021; Ross et al., 2024). Passive acoustic monitoring is now well-poised to build on its successes in tracking ecosystem health and biodiversity to become a pillar of resilience and tipping point monitoring.

In this review, we outline the prospects for passive acoustic monitoring of ecosystem tipping points from a theoretical and practical point of view, including considerations for identifying EWS in acoustic data. Throughout, we focus primarily on critical slowing down-based EWS, given their ongoing dominance in tipping point research, but consider others where appropriate. We close by discussing the key challenges and opportunities for using passive acoustic monitoring to identify EWS.

## Box 1. Basic Theory of Bifurcation Tipping Points

Dynamical systems are characterised by a state, $x$, that changes over time according to a growth function $f(x, B)$ dependent on $x$'s current value and some parameter(s) $B$:

$$\frac{dx}{dt} = f(x, B)$$

Real natural systems are often modelled as stochastic (with a noise term) to capture their continual fluctuation (see Supplementary Material). Classic ecological examples of $x$ are species abundance (population dynamics) or vegetation biomass.

The system will often possess equilibria (steady states)-- values of $x = x^*$ such that $f(x^*, B) = 0$. In single-variable systems, these equilibria are stable states (attractors) when λ, the local gradient ($\frac{df(x,B)}{dx}\Big|_{x=x^*}$) is negative, corresponding to restorative negative feedback (Fig 1c,d). The magnitude of λ is inversely proportional to the system's recovery time from small perturbations (Wissel, 1984). In higher dimensions, stability is determined using the real part of the dominant eigenvalue (see Supplementary Table 1).

Various bifurcation normal forms common to natural systems are brought about by gradual forcing of parameter(s) $B$ that causes λ to tend to zero (Fig. 1c,d). Consequently, stability is progressively lost and the system recovers more slowly from perturbations – so-called 'critical slowing down'. This results in the system displaying longer and larger fluctuations about its current attractor state, such that both autocorrelation (the system's 'memory') and variance increase prior to the bifurcation point, at which λ=0 (derivations in Supplementary Material).

While resilience has been much debated in ecology, Holling's definitions of 'engineering resilience' – the recovery rate from perturbation – and 'ecological resilience' – the amount of disturbance a system can withstand without changing its function – remain prevalent (Holling, 1973, 1996). Both are captured by the potential landscape of the system:

$$V(x) = -\int f(x, B)\, dx$$

Here, stable states are valleys, often termed 'basins of attraction' (Lewontin, 1969), separated by peaks from unstable equilibria (Fig. 1g). In the popular 'ball and cup' analogy of resilience, the system is a ball rolling inside a cup corresponding to the system's current basin of attraction. The ball is always moving due to noise and perturbations, but typically rolls back to the bottom at a rate depending on the steepness of the cup's walls. The steeper, deeper and wider the cup, the more 'resilient' the system.

As $V''(x) = -\lambda$, the cup becomes gentler and shallower as λ approaches zero. Noise or typical perturbations can thus more easily push the system to an alternative attractor (Fig. 1g), should one exist. Hence, while ecological resilience may depend on more than just the potential landscape (e.g., ecosystem reorganisation), both engineering and ecological resilience are linked given their dependence on $\lambda$.

Which bifurcation normal form a system will undergo depends on the nature of the system dynamics. Figure 1 illustrates two widespread cases in ecology using the archetypal example of a single-species population being harvested at an increasing rate (further explained in May (1977) and Clements and Ozgul (2018)): the gradual shift to an alternative stable state (here, extinction) via a transcritical bifurcation (left column), and the sudden, discontinuous jump to an alternative stable attractor via a fold (a.k.a. 'saddle-node' or 'catastrophic') bifurcation, which typifies tipping points with a large, abrupt change that is hard to reverse due to hysteresis.

These dynamics might also manifest in proxies or correlates of the system state. Consider the Asian songbird crisis (the trade of caged songbirds; (Lees and Yuda, 2022)). We might assume that the songbirds collectively undergo logistic population growth and their harvesting saturates with the captured songbird market and/or due to law enforcement restricting poaching, producing similar dynamics to Fig. 1b,d,f. The relative population abundance and dynamics could be inferred from the number of bird vocalisations, detected via a bird call classifier like BirdNET. Further, one might expect the dynamics of Asian songbirds to manifest in the overall soundscape of Southeast Asian forests, which could be captured by relevant soundscape summary statistics, such as the Acoustic Complexity Index, the overall soundscape level, saturation and activity at different frequencies, or even certain machine learning model embeddings (see Section 2.3).

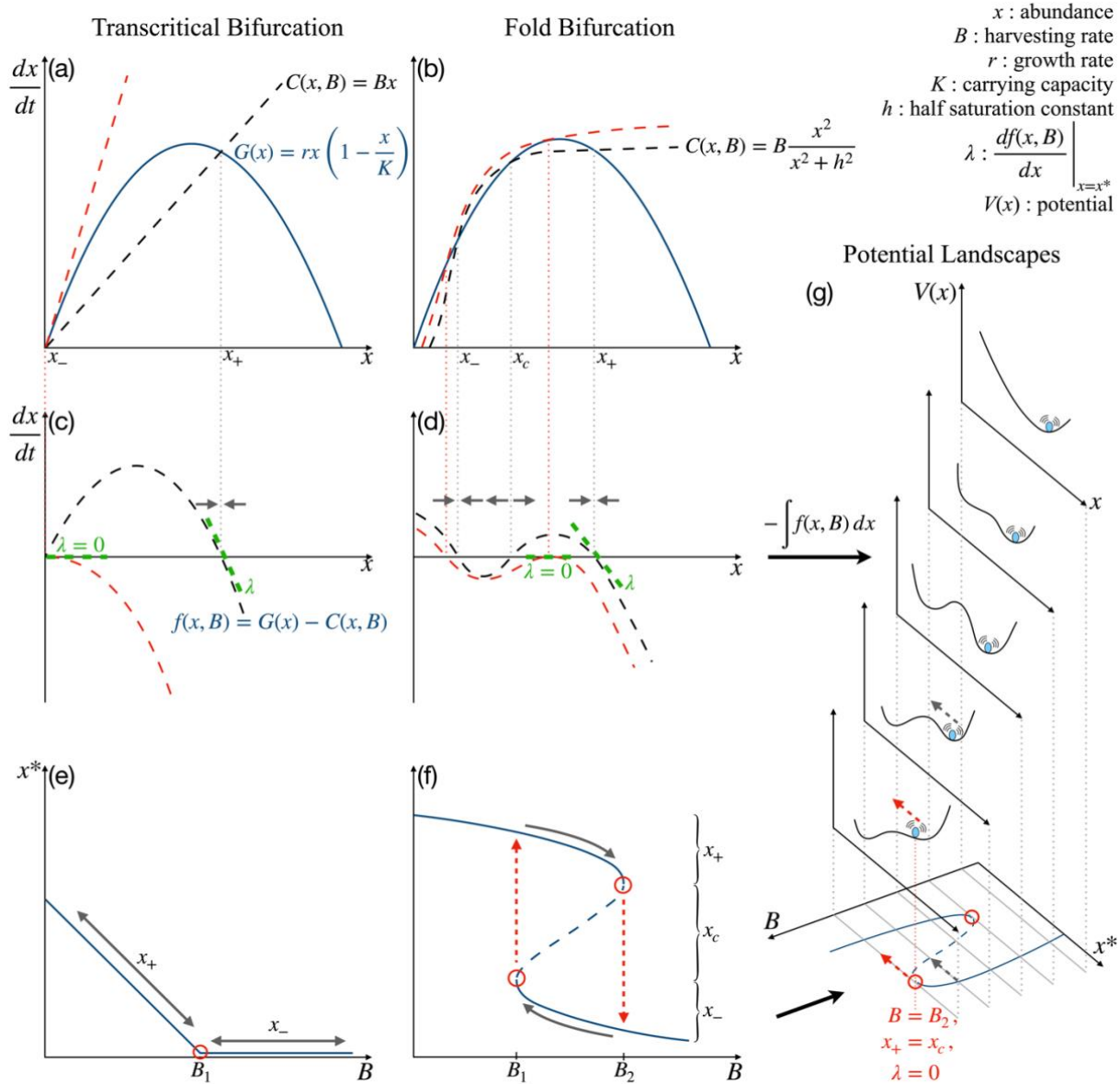

**Figure 1.** Graphs depicting the transcritical and fold bifurcation, illustrated with a single-species population model with harvesting (details in May (1977) and Clements and Ozgul (2018)). Panels **a** and **b** show the population growth rate function $G(x)$ (blue line) and harvesting rate functions $C(x, B)$ (dashed lines). In panel **a** the harvesting increases linearly with abundance $x$, whereas in panel **b** it saturates. Black dashed lines depict a lower harvesting rate, $B$, resulting in multiple equilibria (net zero growth rate) where the harvesting and growth rate functions intersect. Red dashed lines depict harvesting at a higher critical rate at which certain equilibria disappear – the bifurcation point. Panels **c** and **d** show the net growth functions, $f(x, B)$, colour-matched to the two harvesting scenarios in panels **a** and **b**. The net growth functions possess equilibria where these curves cross the $x$-axis at $x_-$, $x_c$ and $x_+$. These equilibria are stable when their gradient, $\lambda$, is negative (corresponding to negative feedback). At the bifurcation point (red dashed lines): in panel **c**, the population has been gradually driven to extinction and the point at zero abundance is now a stable equilibrium; in panel **d**, the stable equilibrium $x_+$ and unstable equilibrium $x_c$ collide and the system jumps to its lower stable equilibrium $x_-$. In both cases, $\lambda$ tends to zero (critical slowing down, see Box 1). Panels **e** and **f** show bifurcation diagrams, in which arrows indicate how the equilibrium states of the systems vary with the parameter $B$ (here, the harvesting rate). Red circles mark bifurcation points and red arrows indicate critical transitions. Note a large decrease in $B$ is required to bring $x_-^*$ back to $x_+^*$ due to hysteresis. Panel **g** illustrates potential landscapes (see Box 1) at different values of $B$ along the fold bifurcation diagram (adapted from (Scheffer et al., 2001)), in which peaks represent unstable equilibria and valleys stable equilibria. Following the 'ball and cup' analogy (Box 1), the system is depicted as a ball constantly moving from noise and perturbations in its current valley (a.k.a. basin of attraction), which gets shallower as λ approaches zero, corresponding to a loss of resilience (Box 1). The arrows here indicate the system moving to its alternate attractor due to the bifurcation itself (red arrows) and sufficiently large noise or a perturbation (grey arrows).

# 2. Step-by-Step Considerations for Acoustically Identifying Tipping Points

## 2.1. Step 1: Will it tip? Identifying tipping potential

While critical slowing down has proven an effective basis for warning of tipping points in a range of systems (Dakos et al., 2024; Scheffer et al., 2009), it can occur without a tipping point and certain tipping points can occur without critical slowing down (Boettiger et al., 2013; Clements and Ozgul, 2018; Kéfi et al., 2013). Thus, determining whether the study system has alternative states that it may tip between is a key first step (Boettiger et al., 2013; Lenton et al., 2024). Understanding the type of transition that can occur, its underlying mechanics, and which variables in a multivariate system are likely to tip and/or show critical slowing down can inform an appropriate strategy for identifying EWS. Although doing so can be challenging and may bias the perceived effectiveness of EWS (Dakos et al., 2024), it is necessary to avoid false alarms, especially given the novelty of using acoustics to anticipate tipping points.

Whether and how an ecosystem may tip can be uncovered in various ways. Theory, models and an understanding of the various feedbacks that maintain stability or drive state transitions provide one basis. For instance, removing a top predator may cause populations at lower trophic levels to explode (Scheffer et al., 2001) while adding predators may drastically alter food web structure (Carpenter et al., 2011). Alternatively, rainforest dieback or deforestation may cause drier local conditions that spur fires and reduce moisture recycling driving further dieback, ultimately tipping to a grassy or treeless state (Staal et al., 2018; Xu et al., 2020). Historical records (e.g., paleoclimate data; (Dakos et al., 2008)) and/or other sensing modalities (e.g., satellite data (Lenton et al., 2024)) may also indicate EWS or evidence of previous regime shifts in the study system (or similar systems). For example, decades of manual point counts were recently used to reveal a tipping point in an at-risk bird community (Almaraz and Green, 2024).

Where existing data reveal alternative stable states across space or time, 'potential analysis' can empirically estimate the potential landscape based on the probability distribution of the system state in relation to a key driving parameter (e.g., temperature or precipitation) (Hirota et al., 2011; Livina et al., 2010). This method has been applied to tropical forests, where remotely sensed tree-cover estimates revealed precipitation thresholds for bifurcations to savannah or treeless states (Hirota et al., 2011). If the study system is impacted by other established tipping systems, tipping in those systems could potentially propagate (Boettiger et al., 2013; Lenton et al., 2024) or cascade (Rocha et al., 2018) into the study system. Whether the study system would show EWS, however, depends how the systems interact (Dakos et al., 2024).

Understanding multivariate systems is more challenging, as it is often unclear how variables will respond to forcing. EWS may be inconsistent across variables and critical slowing down could be less pronounced due to the dominant eigenvector's diminishing importance in a high-dimensional system with many eigenvectors (Boerlijst et al., 2013; Clements and Ozgul, 2018; Dakos et al., 2024). However, various approaches to anticipating transitions in multivariate systems have emerged, such as network analysis and dimensionality reduction techniques (Dakos et al., 2024). Thus, prior to acoustically detecting EWS in a multivariate system, it might be especially important to leverage existing insights into whether it can tip and which key variables or indicator species (Dakos, 2018) might best reflect that possibility.

## 2.2. Step 2: Gathering Data at Sufficient Scale and Resolution

Bifurcations are predicated on the gradual, predominantly monotonic external forcing of the system on a slower timescale than its inherent response to perturbations (Dakos et al., 2024). Detecting critical slowing down requires monitoring long enough to capture forcing while sampling the system state more frequently than its response rate (Lenton et al., 2024). For example, Amazon rainforest vegetation was estimated to have a characteristic recovery time of approximately 1.2 months (Boulton et al., 2022; this arises from the time for leaf regrowth) and faces forcing from human activity and climate change (declining precipitation) on the scale of years to decades. Hence, Boulton et al. (2022) assessed EWS of Amazonian rainforest with over twenty years' remote-sensed vegetation data of monthly temporal resolution.

Theory and such existing data can indicate the required monitoring duration and resolution (e.g., (Smith et al., 2022)). Passive acoustic monitoring offers the advantage of very high (sub-millisecond) temporal resolution and ARUs can record over years (e.g., (Rice et al., 2021; Wrege et al., 2017)), though decadal-scale passive acoustic monitoring is currently rare. Space-for-time substitution may allow for shorter timescales by monitoring along spatial gradients to forecast the fate of a progressively stressed ecosystem (Ciemer et al., 2019; de Oliveira Roque et al., 2018; Eby et al., 2017); monitoring along land-use or disturbance gradients is already relatively common in passive acoustic monitoring (e.g., (Deichmann et al., 2017; Ross et al., 2024; Sethi et al., 2020a; Yoh et al., 2024)).

The required spatial extent of monitoring depends on the scale of the study ecosystem. Autonomous recording units typically detect sounds up to 10s to 100s of meters away, though this strongly depends on sources' sound pressure level and environmental factors (e.g., weather, vegetation density) (Winiarska et al., 2024). Thus, while the full extent of vast ecosystems of up to sub-continental scale, like the Amazon, can be readily monitored by satellites, doing so with acoustics presents a formidable challenge due to the cost and effort of such large deployments. However, large-scale passive acoustic monitoring deployments are feasible (e.g., (Roe et al., 2021; Sethi et al., 2021; Wood et al., 2019)) and can even be achieved through citizen science (e.g., (Mann et al., 2025)). Where only smaller deployments are possible, identifying representative subregions offers one solution, including relevant spatial pressure/disturbance gradients. Monitoring replicates of such gradients or multiple sites at each pressure level can validate the robustness of such methods.

## 2.3. Step 3: Extracting relevant ecological metrics from acoustic data

Methods for extracting ecological information from acoustic data are numerous and may capture ecosystem state variable(s) to monitor for EWS. Most utilise the spectral properties of sound – the amount of energy at different sonic frequencies – usually calculated by applying a Fourier transform to the time-domain acoustic signal (Gibb et al., 2019; Sugai et al., 2019). The frequency distribution is typically represented with spectrograms, 2D image-like plots of the amplitude (colour value/intensity) of different frequency bands (y-axis) over time (x-axis). Broadly, different taxa typically vocalise or otherwise produce sound in specific and sometimes distinct frequency ranges (Krause, 1993; Pijanowski et al., 2011), and different species' calls possess unique spectral and temporal signatures, allowing them to be identified by various means, from manual inspection and simple thresholding to cross-correlation and more advanced machine learning classification (Digby et al., 2013; Gibb et al., 2019; Stowell, 2022; Sugai et al., 2019). Spectral properties are also often inspected to verify the performance of downstream analyses and tools (e.g., (Digby et al., 2013; Le Penru et al., 2025)) as part of so-called 'sound-truthing' (Bradfer-Lawrence et al., 2023; Holgate et al., 2021).

While manual methods were once dominant, the past decade has seen a drive towards faster automated techniques (Digby et al., 2013; Gibb et al., 2019; Sugai et al., 2019), namely acoustic indices, species call classifiers and machine learning model embeddings. These methods will be fundamental to efficiently derive ecological information at the scale required for identifying EWS.

Acoustic indices are 'human-derived' numerical summaries of acoustic data designed to characterise the biophony (sound of organisms), anthropophony (human sound) and geophony (geophysical / environmental sound, e.g., wind and rain) in soundscapes based on various spectral and temporal features (Bradfer-Lawrence et al., 2024, 2023; Sueur et al., 2014). To-date, over 60 have been developed (Buxton et al., 2018b). A prevalent example is the Acoustic Complexity Index, which considers the amount of variability in different frequency bands as a proxy for avian vocal activity or other irregular biophony and geophony (Pieretti et al., 2011). Another, the Normalised Difference Soundscape Index, evaluates the ratio between frequency bands associated with biophony and anthropophony (Bradfer-Lawrence et al., 2024; Sueur et al., 2014). Such indices can describe soundscapes and serve as proxies for biodiversity (Bradfer-Lawrence et al., 2023; Sugai et al., 2019), though recent work has cautioned their use for the latter due to inconsistencies in their relationship to biodiversity between regions and taxa (Alcocer et al., 2022; Sethi et al., 2023). Hence, careful study design and local validation to guide the interpretation of acoustic indices is advised (Bradfer-Lawrence et al., 2023; Metcalf et al., 2021).

Species call classifiers use the time-frequency characteristics of different species' calls to identify vocalising species (Gibb et al., 2019; Stowell, 2022), offering a way to directly measure biodiversity and track the presence, distribution, and behaviour of particular species (e.g., through occupancy models (Wood et al., 2019; Wood and Peery, 2022)). Current state-of-the-art classifiers use a range of different deep learning (DL) models and architectures, in particular various forms of convolutional neural network (CNNs), be they custom-designed or pre-existing architectures such as ResNet (He et al., 2016), VGG (Simonyan and Zisserman, 2015), VGGish (Hershey et al., 2017), and EfficientNet (Tan and Le, 2019). Deep learning classifiers have been developed for many taxa, including birds, cetaceans, bats, elephants, primates, frogs, fish, and insects (examples in Stowell, 2022). Classifiers may be species-specific or able to identify a range of species within one or more taxa, and are typically trained on large datasets (often at least thousands of example calls) under a supervised approach, requiring the often-laborious labelling of training data (though see (Guerrero et al., 2023) on unsupervised approaches). In recent years there have thus been efforts to consolidate datasets and models towards capable, 'off-the-shelf' classifiers that are easy to access and use. Examples include the BTO Acoustic Pipeline, which brings together classifiers for various birds, small mammals, insects and bats (e.g., (Newson and Crisp, 2024)), and BirdNET (Kahl et al., 2021), a widely-used tool now capable of classifying over 6,000 avian species globally, unlocking insights into avian species and communities around the world (Pérez-Granados, 2023; Sethi et al., 2024).

As well as facilitating many ecological studies, pre-existing models like BirdNET are also increasingly being retrained to detect further species/taxa. This approach, termed 'transfer learning' (van Merriënboer et al., 2024), leverages the existing learning of models to learn new classes efficiently, sometimes with very little training data (e.g. < 10 examples) – so-called 'few-shot learning' – hence it is particularly pertinent for rare/endangered species with scarce recordings (Ghani et al., 2023; Nolasco et al., 2023). Transfer learning often leverages existing models' learning by using deep learning model embeddings (Ghani et al., 2023; van Merriënboer et al., 2024). These capture salient features of the input data and are obtained by passing audio into a pretrained deep learning classifier and taking the output of an intermediate (usually, penultimate) model layer. This output is a vector of sometimes hundreds or thousands of numbers that characterises the input recording (sometimes termed a 'feature set' (Sethi et al., 2020b) or 'fingerprint' (Heath et al., 2021)). The embedding vector can be thought of as an abstract, machine-learned, multivariate acoustic index that can describe ecologically-relevant patterns in a soundscape and potentially provide proxies for biodiversity (Ghani et al., 2023; Loo et al., 2025; Sethi et al., 2020b, 2023). As with traditional, human-derived acoustic indices though, model embeddings' relationships to biodiversity can be inconsistent across recording conditions (different regions, recording hardware, etc.), hence they too should be used with caution and careful calibration (Sethi et al., 2023).

Nonetheless, embeddings from the general-purpose audio classifier VGGish (Hershey et al., 2017) have outperformed acoustic indices in various downstream habitat quality and biodiversity classification tasks (Heath et al., 2021; Sethi et al., 2020b), are more robust than acoustic indices to audio compression (Heath et al., 2021), and have supported various other biodiversity assessments (e.g., (Sethi et al., 2022a; Tolkova et al., 2021)). Performance in various ecoacoustic tasks, including few-shot learning is even better in newer general audio classifiers such as YAMNet (CNN) and AudioMAE (self-supervised transformer) and best in embeddings from models pretrained for global birdsong classification, such as BirdNET and Google Perch (Ghani et al., 2023). Loo et al. (2025) also recently demonstrated that training a deep learning model to predict biologically-relevant information inherent to recording metadata – specifically, time of day – could encode ecological information in the model's embeddings, demonstrating prospects for supervised deep learning approaches that avoid manual labelling.

Collectively, these automated methods facilitate the acoustic monitoring of ecosystems from individual species to higher-level biodiversity and ecosystem health metrics across time and space at scale and pace, offering fertile ground for acoustically identifying tipping point EWS. While the system state might not be monitored directly by such methods, they may encode it or provide suitable temporal and spatial proxies, just as satellite sensor products and indices, such as Vegetation Optical Depth bands and the Normalised Difference Vegetation Index, provide proxies for biomass in vegetation resilience-sensing studies (Boulton et al., 2022; Lenton et al., 2022; Smith et al., 2022).

## 2.4. Step 4: Data Pre-Processing

Pre-processing is necessary to verify, clean-up and prepare acoustic monitoring data for calculating EWS. Passive acoustic monitoring data will invariably contain data gaps from barriers to manual data collection (e.g., storms, harsh field conditions) or network outages and insufficient power for fully-autonomous deployments (e.g., solar panels receiving limited sunlight caused by extended cloud cover or obstruction from plant growth or snow). Therefore, down sampling or other methods of data aggregation/averaging, like sliding temporal windows, may be necessary, assuming this does not compromise the required temporal or spatial monitoring resolution. If resolution is compromised, data interpolation or imputation may be required, given data gaps can bias temporal EWS including variance and autocorrelation (De Keersmaecker et al., 2014; Lenton et al., 2024). Interpolation or imputation methods should ideally preserve key statistical properties of the data, such as the overall or seasonal/diel mean, variance and autocorrelation, given these may influence or be directly used for the calculation of EWS. Pre-processing to overcome imperfect data collection is commonplace for other sensing modalities, including satellite remote sensing, where calibration and other operations are necessary due to instrument degradation, cloud cover, snow, shadows and other issues (De Keersmaecker et al., 2014; Lenton et al., 2024). However, while remote sensing can overcome data gaps by using multiple sensors (Lenton et al., 2024), this is more challenging with acoustics' smaller spatiotemporal coverage, underscoring the benefits of monitoring replicate sites.

We also recommend inspecting acoustically-derived ecological metrics and raw audio to ensure that these are as anticipated for the study system's current stable state. For example: do acoustic indices capture expected spatial and temporal trends, such as peaks in avian activity during dawn and dusk choruses? Do deep learning embeddings follow similar distributions between site or season replicates? If patterns are unexpected, inspecting recordings may reveal underlying issues, such as self-noise or microphone degradation (Bradfer-Lawrence et al., 2023). Such issues might be rectified by applying appropriate frequency filters to remove undesired noise. It is also common practice – and recommended (Knight et al., 2017; Wood and Kahl, 2024) – to calibrate prediction score thresholds in species call classifiers to particular ecosystems and recording hardware by working with experts to manually label calls in a subset of recordings as a form of ground-truth (Le Penru et al., 2025; Sethi et al., 2024). Thresholds that optimise performance (usually a trade-off between precision and recall – ensuring predictions are accurate and capture a large proportion of calls) can then be set (Sethi et al., 2024) and/or species-specific probability thresholds derived (Wood and Kahl, 2024).

Many temporal EWS, including variance, autocorrelation and certain deep learning methods, require detrending of the system state (or state proxy) time series to separate long-term or seasonal trends from the short-term fluctuations in which critical slowing down occurs (Dakos et al., 2012a). The fluctuations are usually obtained by taking the residuals of the detrending process. Detrending methods include simple subtraction of the long term mean (Held and Kleinen, 2004; Lenton et al., 2012), Gaussian filtering (Dakos et al., 2008; Lenton et al., 2012), and Lowess smoothing (Boulton et al., 2022; Bury et al., 2021). The nature of de-trending can impact EWS (Lenton et al., 2012). For instance, Dablander and Bury (2022) revealed optimal accuracy from a deep learning classifier for predicting bifurcations in time series data requires using the detrending method(s) applied to the model's training data. In the case of spatial EWS, pre-processing can also be required, such as applying a discrete Fourier transform to spatial data (Kéfi et al., 2014) or to assemble data gathered along spatial gradients in a space-for-time substitution (e.g., (de Oliveira Roque et al., 2018; Eby et al., 2017)).

## 2.5. Step 5: Searching for Early Warning Signals

The classic, most widespread temporal EWS of critical slowing down are variance and autocorrelation (Dakos et al., 2024). These indicators are usually calculated on sliding windows that increment at a suitable resolution to capture system state fluctuations. The choice of window size can affect results, with longer windows providing smoother and more reliable EWS (Lenton et al., 2012). However, it is important to note that for each increment of the window, the time series should be approximately stationary (i.e., the statistical properties, such as the mean, variance and autocorrelation do not vary over time). It is advisable to try a reasonable range of sliding window lengths to verify whether the calculated EWS are robust to window size (e.g., (Boulton et al., 2022)). Spurious trends in variance and autocorrelation unrelated to critical slowing down can arise. For example, variance is more susceptible to large perturbations or shocks and may be more likely than autocorrelation to increase due to 'one-off' events (e.g., a once-in-a-century drought) (Boulton et al., 2022). However, variance may also decrease near a bifurcation due to limited data or a reduction in external noise before tipping (Dakos et al., 2012b). Autocorrelation is therefore generally thought to be a more robust indicator, though it can be influenced by changes to noise in the data. For instance, mixing newer and older satellite sensors for greater spatiotemporal coverage may increase autocorrelation, since uncorrelated measurement noise decreases as newer, less noisy sensors are introduced (Smith et al., 2023). Therefore, it is recommended to ensure both variance and autocorrelation exhibit similar trends for robust results (e.g., Boulton et al., 2022).

Other temporal EWS include further statistical changes, such as spectral reddening (a shift in the power spectrum towards lower frequencies) (Biggs et al., 2009) and an increase in skewness towards the system's prevailing alternative state (Guttal and Jayaprakash, 2008). Under strong stochastic forcing, the system may also 'flicker', i.e., make brief jumps to an alternate attractor (Dakos et al., 2024). Skewness, variance and autocorrelation also have spatial equivalents (Dakos et al., 2010; Guttal and Jayaprakash, 2009) and complement other, more complex spatial EWS such as Turing patterns (increasing self-organisation) and Kolmogorov complexity (Dakos et al., 2024; Kéfi et al., 2014), though it is less clear whether or how these could be captured acoustically. Further temporal critical slowing down-based methods include fitting statistical models, e.g., detrended fluctuation analysis (Livina and Lenton, 2007), or using process-based models (Dakos et al., 2024). Deep learning models for time series classification also show potential. Bury et al. (2021) developed a deep learning model that outperforms traditional EWS at predicting bifurcations on a variety of model and empirical systems, while also predicting the normal form associated with the bifurcation (fold, transcritical or Hopf). This can offer vital insight into the nature of the transition and inform appropriate management strategies.

Another highly promising class of EWS that acoustics is well-poised to monitor are trait-based signals (Clements and Ozgul, 2018). There is notable theoretical and empirical evidence that fitness-related, plastic phenotypic traits – namely, body size – may shift as environmental conditions change (Clements and Ozgul, 2018). In microcosm studies, Clements and Ozgul (2016) showed that the mean body size of a stressed population declined on a significant scale ahead of a collapse to extinction, while the population abundance displayed conventional EWS including rising autocorrelation. Further work demonstrated similar trait and abundance signals in the past collapse of overharvested whale populations (Clements et al., 2017). Acoustics may be able to detect such declines in body size in vocal animals through changes to the nature of vocalisations (Bowling et al., 2017; Martin et al., 2017). Species call classifiers could, for instance, both identify the number of calls from different species, offering insight into relative abundance dynamics, and parse calls for analyses relating acoustic features (e.g., amplitude, fundamental frequency, nature of harmonics) to body size.

Overall, the development and observation of EWS in natural systems has proliferated in the past two decades. We have highlighted certain prominent, acoustics-relevant examples, but broader reviews can be found in (Clements and Ozgul, 2018; Kéfi et al., 2014; Scheffer et al., 2015) and Dakos et al. (2024), who also outline software packages for calculating EWS. We expect that integrating passive acoustic monitoring into the study of EWS may unleash another powerful route to understand and anticipate ecosystem tipping points.

# 3. Opportunities and Challenges for Acoustically Monitoring Tipping Point Early Warning Signals

## 3.1. Acoustics may reveal early warning signals at various ecosystem levels

Ecoacoustics can capture both detailed and high-level information about ecosystems, offering a myriad of routes to monitor salient ecological variables for EWS. Which to prioritise may depend on the specific question at hand and/or outcomes of datamining large ecoacoustic datasets for EWS. Species call classifiers can uncover the presence and activity of vocal species (Ross et al., 2023), and thereby reveal their relative population dynamics and the fluctuations in which critical slowing down manifests. Species abundances can also be directly estimated by using microphone arrays to determine the source direction or location of calls (Blumstein et al., 2011; Heath, 2022; Rhinehart et al., 2020). By distinguishing sounds arriving from different directions at a single point, autonomous recording units with small, co-located microphone arrays (e.g., (Heath et al., 2024)) could potentially even detect certain spatial EWS or facilitate space-for-time substitutions with fewer recorders than sites – for instance by comparing variance and autocorrelation on ecoacoustic signals from neighbouring habitats.

There is a unique opportunity to monitor potential population collapses in terms of phenotypic trait changes that leave acoustic signatures in vocalisations (Bowling et al., 2017; Clements et al., 2017; Clements and Ozgul, 2016; Martin et al., 2017).

Additionally, various soundscape analysis methods have to-date revealed different ecosystem responses to disturbance, including soundscape resistance and recovery (Burivalova et al., 2018; Deichmann et al., 2017; Gasc et al., 2018; Ross et al., 2024; Wood et al., 2024). Gottesman et al. (2021) specifically note that soundscape variability could indicate landscape disturbance and, potentially, impending regime shifts.

Acoustic indices and deep learning model embeddings may also capture the state variable(s) of systems from communities to overall ecosystems, given their correlations with and ability to predict ecosystem health and biodiversity. While acoustic indices may be easier to relate mechanistically to the system state, embeddings' capacity to encode it could be uncovered by attempting to predict the system state variable(s), assuming ground truth state data exists. A pertinent example could be predicting above-ground biomass, as model embeddings from tropical forest recordings have been used to do so before (Sethi et al., 2020b), and satellite-sensed vegetation data has revealed increasing EWS of tipping points in global vegetation (Lenton et al., 2022; Smith et al., 2022), including tropical forests (Boulton et al., 2022; Hirota et al., 2011; Verbesselt et al., 2016). Whether the biomass predictions or EWS derived from them match those from other modalities like satellite imagery could indicate how suitable certain model embeddings are for calculating EWS.

As deep learning embeddings are high-dimensional, they could capture several variables in multivariate systems (e.g., multi-species population dynamics). The question for univariate or multivariate systems is in which embedding feature(s) to look for EWS. One answer could be to use dimensionality-reduction techniques, which are not only used to apply univariate EWS to multivariate systems (Bathiany et al., 2013; Dakos et al., 2024; Held and Kleinen, 2004), but also commonly applied to deep learning embeddings and other high-dimensional data for visualisation and lower-dimensional analyses (e.g., (Bravo Sanchez et al., 2024)). Principal Component Analysis (PCA), which seeks to capture much of the variance and overall directions of change in few new variables ('principal components' that linearly combine the original variables), may be particularly promising (Dakos et al., 2024; Greenacre et al., 2022). PCA has been used in studies aiming to anticipate transitions in climate and biological systems (Bathiany et al., 2013; Dakos, 2018; Dakos et al., 2024; Held and Kleinen, 2004; Lever et al., 2020) and could be applied to deep learning model embeddings in a parallel analysis to capture the overarching dynamics of a complex system. The first principal component could then be assessed for univariate EWS and may suggest which embedding features contribute most to variability in the system based on the size of their loadings (coefficients). Those with the largest loadings could be investigated for EWS and their correlations to biodiversity, to relate results back to the system's biology.

## 3.2. Acoustic data is limited but growing

This represents both a challenge and an opportunity. On one hand, acoustic monitoring has to-date been of limited spatiotemporal scale compared to the large coverage needed to detect EWS in many systems. Longer passive acoustic monitoring efforts are currently closer to years in scale than the decades needed to capture trends in forcing from human pressures and climate change. On the other hand, this is changing with the falling cost and increasing capability of tools for each stage of acoustic monitoring, from recorder hardware to data storage and cloud computing (e.g., (Sethi et al., 2020a)). The advent of cheaper, robust autonomous recording units (e.g.,

AudioMoth (Hill et al., 2018, 2019), Cornell's Swift/SwiftOne (Koch et al., 2016), BUGG (bugg.xzy) and its DIY predecessor (Sethi et al., 2018)) has been instrumental in facilitating longer and larger deployments. Power efficiency gains (partly thanks to temporally sub-setting recording) allow battery-powered units to now last for months before requiring manual battery replacement (e.g. (Hill et al., 2018)), while advances in fully-autonomous recorders, with self-sufficient powering and data transmission (Heath et al., 2024; Sethi et al., 2018), can unlock truly long-term, near real-time monitoring that minimises manual intervention post-deployment. Drone-deployed biodegradable sensors may increase the scale of future monitoring even further – especially in hard-to-access areas – while minimising environmental impacts (Sethi et al., 2022b). In the marine realm, the recent discovery that whale vocalisations can be detected with fibre-optic telecommunication cables has potentially uncovered a global-scale, high resolution, real-time system for monitoring ocean health and biodiversity (Bouffaut et al., 2022).

With the growing number of tools and datasets, however, comes the risk of disparate methods hindering collaboration and data assimilation for greater spatiotemporal coverage. Ecoacoustic metrics can be easily misinterpreted without study-specific calibration (Bradfer-Lawrence et al., 2023; Sethi et al., 2023; Wood and Kahl, 2024) and EWS can be impacted by combining data from different sensors, such as inconsistent measurement noise biassing autocorrelation (Smith et al., 2023). Further, while ecoacoustic monitoring tools are becoming less expensive, they remain costly especially for researchers based in less wealthy nations, where even costs associated with storing large passive acoustic monitoring datasets can be significant. To deal with this, raw data may need to be compressed or deleted after extracting relevant metrics, limiting future reuse of such datasets in large-scale collaborative monitoring studies.

These challenges can partly be alleviated by harmonising methods with the growing body of ecoacoustic monitoring guidelines. These include guides on compression and recording schedule (Heath et al., 2021), acoustic indices (Bradfer-Lawrence et al., 2024, 2023), species call classifiers (Wood and Kahl, 2024), and overall monitoring pipelines (Metcalf et al., 2023; Teixeira et al., 2024). Further, there are now various efforts to track global passive acoustic monitoring projects and assimilate their metadata, such as the Worldwide Soundscapes project (Darras et al., 2025). In addition, patchy or small-scale data could be complemented by data from other monitoring modalities, such as remote sensing, camera traps, environmental DNA, LiDAR, and more (Besson et al., 2022). There has in recent years been increasing work on multimodal environmental and biodiversity sensing (e.g., (Buxton et al., 2018a; Gottwald et al., 2021; Wägele et al., 2022)) including pairing satellite imagery and high-throughput automated recorders (Bush et al., 2017). With this, there is a growing need for multimodal EWS that are more robust to data gaps, which could potentially be met by multimodal AI. There is also a notable opportunity to integrate multimodal monitoring data into more complex holistic ecosystem models, such as the nascent Virtual Ecosystem (Ewers et al., 2024). Such models could be used to simulate state transitions and/or provide high-resolution synthetic system state data on large spatiotemporal scales, with regular calibration from live monitoring, for calculating multitudinous EWS.

## 3.3. Uncertainties in tipping point theory and early warning signals

Ecosystems and their resilience are complex (Dakos and Kéfi, 2022; Holling, 1973; Pimm, 1984; Scheffer et al., 2015; Walker, 2020; Walker et al., 2004). Many ecosystems possess cyclic (Scheffer et al., 2015), transient (Hastings, 2004), or chaotic dynamics (Hastings et al., 1993; May et al., 1997) that may not display typical critical slowing down-based EWS or fold/transcritical bifurcations (Boettiger et al., 2013; Kéfi et al., 2013) and the slow, broadly monotonic forcing they assume. Further, many understandings of ecosystem resilience might not conform to the tractable mathematics of lower-dimensional dynamical systems and bifurcation normal forms. Ecosystem resilience can encompass other ways of adapting to persist in function when faced with disturbance and changing conditions (Oliver et al., 2015; Sánchez-Pinillos et al., 2024; Walker, 2020). In such cases, EWS not based on critical slowing down may be able to anticipate tipping points, but better models and quantitative metrics of the other mechanisms of ecological resilience are needed (e.g., (Dakos and Kéfi, 2022; Sánchez-Pinillos et al., 2024)). Nonetheless, critical slowing down offers a strong basis for forewarning of tipping points for a remarkably wide range of real natural systems, including some with cyclic dynamics (Scheffer et al., 2015), hence it remains fundamental to assessing ecosystem resilience.

From a practical perspective, each step of monitoring EWS may introduce potential uncertainties about the accuracy and reliability of results, from data gathering challenges (e.g., recorder self-noise, data gaps) that may hamper downstream analyses to the extraction of ecological metrics, data pre-processing, and EWS calculation, all of which are sensitive to analysis parameters. Such uncertainties are particularly pronounced with the use of 'black box' deep learning tools, as the rationale for their predictions and 'meaning' of their embeddings are not fully known (but can be explored with correlation analyses). The abstraction of these tools may call into question whether their combined outputs, such as applying a bifurcation classifier to a time series of the first principal component from acoustic deep learning embeddings, have any real meaning or biological basis.

However, many of these uncertainties can – and have – been investigated (Dablander and Bury, 2022; Lenton et al., 2012) and overcome for other sensing modalities including remote sensing (Boulton et al., 2022; Lenton et al., 2024; Smith et al., 2023), leaving a clear roadmap for doing so with acoustic data. That includes grounding tipping point forecasting in a rigorous ecological understanding of the system and its capacity to tip alongside comprehensively robustness testing results and checking for consistency with EWS from other sensing modalities. With such careful study design and validation, passive acoustic monitoring could uncover pertinent EWS across scales.

# Conclusion

As anthropogenic climate change and land use further intensify, the threat of numerous climate, ecological and Earth system tipping points is multiplying, with dire and potentially cascading consequences. To mitigate such consequences and even trigger positive tipping points that combat them – such as rapid rewilding and decarbonisation – we need tools for effectively anticipating tipping points and studying the health of affected systems more broadly. Acoustic monitoring has proven its capacity for the latter, and shows considerable promise for the former, given its ability to capture ecosystems at multiple levels, exceptionally high resolution and, in the case of fully autonomous recorders, in near real-time. While tipping point theory might not apply to all ecosystems and the scale and compatibility of ecoacoustic deployments are relatively small at present, there is now a crucial opportunity to integrate acoustics into tipping point forecasting systems, owing to the confluence of: (i) increasingly capable, low-cost hardware powering larger acoustic monitoring deployments; (ii) advancing tools and guidelines for the automated extraction of ecological information from soundscape recordings; and, (iii) improving confidence and methods in tipping point early warning signals. Leveraging acoustics to this end may prove vital in detecting the collapse of ecosystems that underpin nature and society just in time to intervene.

## Acknowledgements

We thank Marten Scheffer, Sharon Martinson, Fabian Dablander, Irina Tolkova, Laural Symes, Connor Wood, Wendy Erb and Becky Heath for helpful discussions. We also thank Taran Rallings, David Orme and Hollie Folkard-Tapp for feedback on earlier versions of this manuscript. Neel P. Le Penru was supported by the Natural Environment Research Council [grant number NE/S007415/1] via the Science and Solutions for a Changing Planet DTP. Thomas M. Bury is supported by a Fonds de Recherche du Québec—Nature et Technologies (FRQNT) postdoctoral fellowship.

Supporting Information for *Prospects for Acoustically Monitoring Ecosystem Tipping Points*

# Supporting Information for *Prospects for Acoustically Monitoring Ecosystem Tipping Points*

**Supplementary Table 1.** Expanded glossary of key concepts underpinning tipping point theory, explained in terms of their mathematical and intuitive/ecological definitions.

| Concept | Mathematical Definition | | Intuitive or Ecological Definition |
|---|---|---|---|
| | *In univariate (1D) systems* | *In higher dimensions* | |
| **Dynamical System** | $\dfrac{dx}{dt} = f(x, B)$ | Same but in vector / matrix notation | A system that changes over time in a manner defined by the growth function $f(x, B)$, which depends on the current state of the system, $x$, and the system parameter(s) $B$ (i.e., conditions, such as climate or environmental conditions like precipitation, temperature etc.). |
| **Stochastic Dynamical System** | $dx = f(x, B)dt + g(x, B)dW_t$<br><br>Where:<br>$f$ is the deterministic ('drift') term,<br>$g$ is the stochastic ('diffusion') term. | Same but in vector / matrix notation | System with certain degree of randomness such that it is ever-fluctuating about its current stable state.<br><br>Most real natural systems are understood to be stochastic due to variability in their forcing (changing of the system parameter(s)) or random internal noise (captured here by $dW_t$). |
| **Equilibrium** | A state $x^*$ where $f(x^*, B) = 0$ | | Values of the state variable where the growth function is zero. As a consequence, if the system is at an equilibrium and is unperturbed, it will remain at this equilibrium. |
| **Stability** | Determined by:<br>$\left.\dfrac{df(x, B)}{dx}\right\|_{x=x^*}$<br><br>Gradient of $f(x, B)$ w.r.t. $x$, evaluated at equilibrium $x^*$.<br><br>Can be denoted $f'(x^*, B)$ or, often, $\lambda$.<br><br>$\lambda < 0 \rightarrow$ stable,<br>$\lambda > 0 \rightarrow$ unstable. | Determined by the dominant eigenvalue (eigenvalue with largest real part) of the Jacobian matrix (matrix of first order partial derivatives) of $f(x, B)$. The dominant eigenvalue, also denoted $\lambda$, is a measure of stability along the direction in state space that is least stable.<br><br>Re$\lambda < 0 \rightarrow$ stable,<br>Re$\lambda > 0 \rightarrow$ unstable. | Stability refers to the behaviour of a system following small perturbations from an equilibrium.<br><br>An equilibrium point is said to be stable if trajectories return to equilibrium following perturbation. In contrast, trajectories move away from unstable states when perturbed.<br><br>The dominant eigenvalue, $\lambda$ (which is the local gradient at equilibrium in univariate systems), gives a measure for the rate at which the system returns to an equilibrium (if stable) or moves away from it (if unstable). |

| Concept | Mathematical Definition | | Intuitive or Ecological Definition |
|---|---|---|---|
| | *In univariate (1D) systems* | *In higher dimensions* | |
| **Characteristic Return Time** | $T_R = -1/\lambda$ | $T_R = -1/\text{Re}\lambda$ | Provides a time scale that quantifies how quickly the system returns to equilibrium. |
| | As a system returns to equilibrium following a perturbation, its distance from equilibrium reduces by a factor of *e* after a time $T_R$ (Wissel, 1984). | | |
| **Critical Slowing Down** | The dominant eigenvalue approaching zero from below, due to changing parameters. This corresponds to the characteristic return time approaching infinity. | | System responds more slowly to perturbations following a loss in resilience decreasing system stability, e.g., forest takes longer to recover from drought or fire. |
| **Potential Landscape** | The potential of a system is defined: $$V(x) = -\int f(x, B) dx$$ Note that this will have stationary points (peaks and valleys) when $f(x, B) = 0$, i.e. at *x\**, but the negative sign makes the stable equilibria into valleys and unstable ones into peaks. | | While the concept of potential is somewhat arbitrary in ecological terms (square of system state per unit time), it serves as a useful mathematical and conceptual model for the stability and resilience of a system, with the valleys at stable equilibria corresponding to '**basins of attraction**' (Lewontin, 1969) separated by peaks from unstable equilibria.<br><br>This is also sometimes described in terms of the '**ball-and-cup**' model wherein the system state corresponds to a ball that moves along the potential landscape (with no inertia) and the basins of attraction are the 'cups'. At a stable equilibrium, the ball is at the bottom of one of the cups but is constantly moving from small perturbations due to stochasticity (randomness) in external forcing and random internal noise. However, the walls of the 'cup' cause it to constantly roll back to equilibrium, at a rate depending on the steepness of the walls – which we can see from the mathematics relates to $\lambda$.<br><br>The amount of disturbance that the system can withstand is described by the shape and size of the basin of attraction (Dakos and Kéfi, 2022). Specifically, the width indicates how much the system state variable can be changed before it leaves the current basin of attraction, while the depth and steepness relate to how much the forcing and noise terms can change before the system switches to an alternative attractor. |

| Concept | Mathematical Definition | | Intuitive or Ecological Definition |
|---|---|---|---|
| | *In univariate (1D) systems* | *In higher dimensions* | |
| | If the underlying dynamics of the system are not known, one option to estimate the potential landscape is to integrate the Taylor expansion of the growth function at some small perturbation about equilibrium, $f(x^* + \eta)$ (as in e.g., Bury et al., 2021): $$\frac{d(x^* + \eta)}{dt} = f(x^* + \eta)$$ $$= f(x^*) + \left.\frac{\partial f}{\partial x}\right|_{x^*} \eta + \frac{1}{2}\left.\frac{\partial^2 f}{\partial x^2}\right|_{x^*} \eta^2 + \ldots$$ $$f(x^* + \eta) = \lambda_1 \eta + \lambda_2 \eta^2 + \ldots$$ $$V(\eta) = -\int f(x^* + \eta)d\eta = -\frac{1}{2}\lambda_1 \eta^2 - \frac{1}{3}\lambda_2 \eta^3 + \ldots$$ | | If we only take the Taylor expansion of $f(x^* + \eta)$ to the first-order term (a.k.a. linearising the equation), the resulting potential landscape will only have a quadratic term corresponding to a parabola. This is a reasonable approximation of the potential landscape near the equilibrium (capturing the dynamics for small perturbations) and will get shallower as $\lambda_1$ (the dominant eigenvalue) approaches zero. However, it ignores other stable equilibria, which require higher-order terms. Adding a cubic term will add a 'peak' roughly corresponding to the unstable equilibrium, but a bi-stable system will require at least a fourth-order term. Even with just the cubic approximation, though, as $\lambda_1$ approaches zero the peak and valley approach and eventually coincide, becoming a flat point of inflection leaving the system to fall into a neighbouring basin of attraction. |
| **Resilience** | A wide range of definitions and associated metrics of resilience have been proposed, though the two most prevailing were put forward by Holling (1973, 1996) and are termed **'engineering resilience'** and **'ecological resilience'**. | | |
| | **Engineering resilience** essentially corresponds to stability and the characteristic return time from perturbations, $T_R$. | | How fast a system recovers after perturbation (Holling, 1996, 1973; Pimm, 1984). |
| | **Ecological resilience** is generally characterised by the shape and size of the basin of attraction – a number of metrics related to the features of it and the potential landscape are outlined in (Dakos and Kéfi, 2022). | | The amount of disturbance, perturbation or forcing an ecosystem can endure without changing its function and passing to an alternative stable state. This can be facilitated by mechanisms that are not necessarily fully captured by dynamical systems theory, such as the capacity of ecosystems to reorganise, though such changes should result in an altered potential landscape and basin of attraction. |
| **Bifurcations and Normal Forms** | In dynamical systems theory, a bifurcation is a qualitative, topological change in the dynamics of a system in response to gradual, smooth change in the system parameter(s) – i.e., condition(s) passing a critical threshold (Lenton, 2013; Strogatz, 2024). For bifurcation-induced tipping points, this change is typically the loss of the current equilibrium state's stability, resulting in a transition to an alternate stable state. Bifurcations can be grouped into basic, lower-dimensional 'normal forms' which characterise different responses to forcing and to which the dynamics of high dimensional systems simplify near a bifurcation (**centre manifold theorem**). Three of the most prevalent in ecology are: | | |

| Concept | Mathematical Definition | | Intuitive or Ecological Definition |
|---|---|---|---|
| | *In univariate (1D) systems* | *In higher dimensions* | |
| | **Transcritical bifurcation** – continuous, gradual transition from one stable state to another.<br><br>This corresponds to a 'kink' in the **bifurcation diagram** (which plots the system's equilibrium values against the system parameter(s); Fig. 1e,f Main Text) owing to the second order derivative of the equilibrium values with respect to the system parameter(s) (i.e., of the bifurcation diagram's curve; $\frac{d^2 x^*}{dB^2}$) responding discontinuously to forcing (Clements and Ozgul, 2018; Lenton, 2013). | | This can characterise "the transition to population extinction" (Dakos and Soler-Toscano, 2016, p. 145).<br><br>Some of earliest transcritical bifurcations in ecology were effectively first described by Lotka (1925). |
| | **Hopf bifurcation** – a transition from a single stable equilibrium to "stable cyclical dynamics" (Clements and Ozgul, 2018, p. 906). | | First described in ecology in terms of "the onset of oscillations in predator-prey models" (Dakos and Soler-Toscano, 2016, p. 145) by Rosenzweig and Macarthur (1963). |
| | **Fold** (or sometimes, "catastrophic" or "saddle node") **bifurcation** – characterised by an abrupt and discontinuous shift to an alternative equilibrium (Clements and Ozgul, 2018, p. 906). It is the result of a stable and an unstable equilibrium colliding and annihilating one-another (Fig. 1b,d,f Main Text). | | Fold bifurcations typify tipping points in dynamical systems given they are particularly abrupt and difficult (sometimes impossible) to reverse, owing to **hysteresis**: the need to return to a lower parameter threshold than that which brought about the new state in order to revert to the previous stable state. The name refers to the 'fold' in the curve of the bifurcation diagram (Fig. 1f, Main Text).<br><br>Fold bifurcations were first described in ecology in terms of "resource overexploitation" (Dakos and Soler-Toscano, 2016, p. 145) in Noy-Meir's (1975) extension of graphical analyses of predator-prey models by Rosenzweig and Macarthur's (1963) to grazing systems. They were further discussed in relation to various classic ecological models (including for harvesting animal populations and pest outbreaks) in a seminal review paper by May (1977). |
| **Tipping Points** | The point at which small change can make a big difference to a system (Gladwell, 2000; Lenton et al., 2008). Specifically, where a small perturbation or increment in the system parameter(s) (e.g., climate and environmental conditions) causes a qualitative change to the system driven by strong positive feedback, such as a transition to an alternate stable state.<br><br>Use of the term has varied in some previous literature (van Nes et al., 2016). As it is typified by fold bifurcations (but can describe other phenomena that produce state transitions or other qualitative change), it is sometimes interpreted as being synonymous with the bifurcation point. However, tipping points can be caused by noise or small perturbations near the bifurcation point, which may be sufficient to push the system out of its current basin of attraction to an alternate stable state. | | |

| Concept | Mathematical Definition | Intuitive or Ecological Definition |
|---|---|---|
| | *In univariate (1D) systems* | |
| **Early Warning Signal - Autocorrelation** | In continuous time, the autocorrelation of the system state fluctuations is given by: $$\rho(\tau) = e^{\lambda \tau}$$ It thus tends to one as $\lambda \to 0$ from below. In discrete time, if the system can be well described by an autoregressive process of order 1 (AR(1)) then the lag-1 autocorrelation is equivalent to the AR(1) coefficient: $$\alpha = e^{\lambda \Delta t}$$ Thus the lag-1 autocorrelation and AR(1) coefficient also tend to 1 for processes best modelled in discrete time as $\lambda \to 0$ from below. Full derivations are provided below. Please note that the equations here are for 1D systems and become more complex in higher dimensions. | Critical slowing down causes the system's current state to be more similar to its recent previous states (sometimes described conceptually in terms of rising memory of the system (Clements and Ozgul, 2018; Kéfi et al., 2014)), hence there is an increase in the system state time series of autocorrelation at lag-1. This also holds true for further lags for transcritical and fold bifurcations, but is not necessarily the case for all types of bifurcation, such as Hopf bifurcations, for which autocorrelation may increase or decrease at subsequent lags (Bury et al., 2020). |
| **Early Warning Signal - Variance** | In continuous time, the variance of the system state fluctuations is given in the stationary regime ($t \to \infty$, assuming $\lambda < 0$) by: $$\text{Var}(\eta) = -\frac{\sigma^2}{2\lambda}$$ Where $\eta = x - x^*$. The variance thus tends to infinity as $\lambda \to 0$ from below. In discrete time, the variance is given by: $$\text{Var}(\eta) = \frac{\sigma^2}{1 - \alpha^2}$$ Where $\alpha$ is the AR(1) coefficient introduced in the autocorrelation section above. Since $\alpha \to 1$ as $\lambda \to 0$, the discrete time variance also tends to infinity under critical slowing down. Full derivations are also provided below. As above, these equations are for 1D systems and are more complex in higher dimensions. | As critical slowing down produces longer and larger fluctuations in the system state due to the system's declining capacity to recover from perturbations, variance of the system state and its fluctuations increase over time. |

# 1. Derivation of variance and lag-1 autocorrelation as early warning signals of tipping points

## 1.1. Continuous-Time Process

A general stochastic process with additive white noise may be written

$$\frac{dx}{dt} = f(x) + \sigma \xi(t) \qquad (1)$$

where $f$ describes the deterministic dynamics, $\sigma$ is the noise amplitude, and $\xi(t)$ is a Gaussian white noise process satisfying $\mathbb{E}[\xi(t)] = 0$ and $\mathbb{E}[\xi(t)\xi(s)] = \delta(t-s)$. Assuming a stable equilibrium at $x^*$, and letting $\eta(t) = x - x^*$, we have

$$\frac{d\eta}{dt} = f(x^* + \eta) + \sigma \xi(t), \qquad (2)$$

an equation which describes the dynamics about equilibrium. For sufficiently small deviations from equilibrium, the dynamics are well approximated by low-order terms in the Taylor expansion. Expanding $f(x^* + \eta)$ in a Taylor series about $x^*$ and retaining only the linear term (valid for small $\eta$),

$$\frac{d\eta}{dt} = \lambda \eta + \sigma \xi(t), \qquad (3)$$

where $\lambda = \frac{df}{dx}\big|_{x=x^*}$. This equation is known as an Ornstein Uhlenbeck Process. Its statistical properties are well known (e.g., *Stochastic Methods* (Gardiner, 2009)). It can be solved using an integrating factor $e^{-\lambda t}$, and has solution

$$\eta(t) = \eta(0)e^{\lambda t} + \sigma e^{\lambda t} \int_0^t e^{-\lambda t'} \xi(t') dt'. \qquad (4)$$

In the long-time limit ($t \to \infty$), the contribution from the initial condition $\eta(0)e^{\lambda t}$ vanishes since $\lambda < 0$ (the equilibrium is stable).

### 1.1.1. Variance

The variance is defined as

$$\text{Var}(\eta) = \mathbb{E}[(\eta - \mathbb{E}[\eta])^2] \qquad (5)$$

Since $\mathbb{E}[\xi(t)] = 0$ and the process is linear in $\xi$, we have $\mathbb{E}[\eta(t)] = 0$, so

$$\text{Var}(\eta) = \mathbb{E}[\eta^2] \qquad (6)$$

Substituting the solution for $\eta(t)$ in the long-time regime (where the deterministic term vanishes), we get

$$\text{Var}(\eta) = \mathbb{E}\left[\left(\sigma e^{\lambda t} \int_0^t e^{-\lambda t'} \xi(t'), dt'\right)^2\right] \qquad (7)$$

$$= \sigma^2 e^{2\lambda t} \int_0^t dt' \int_0^t dt'', e^{-\lambda(t'+t'')} \mathbb{E}[\xi(t')\xi(t'')]. \qquad (8)$$

Using the white noise property $\mathbb{E}[\xi(t')\xi(t'')] = \delta(t' - t'')$, we find

$$\text{Var}(\eta) = \sigma^2 e^{2\lambda t} \int_0^t dt', e^{-2\lambda t'} \tag{9}$$

$$= \sigma^2 e^{2\lambda t} \cdot \frac{1 - e^{-2\lambda t}}{2\lambda} \tag{10}$$

$$= \frac{\sigma^2}{2\lambda} \left(e^{2\lambda t} - 1\right), \tag{11}$$

which simplifies in the stationary regime ($t \to \infty$, assuming $\lambda < 0$) to

$$\text{Var}(\eta) = -\frac{\sigma^2}{2\lambda}. \tag{12}$$

Note that as $\lambda \to 0$ from below, the variance diverges, indicating a loss of stability near bifurcation (i.e., critical slowing down). This is the equation given in Dakos and Kéfi (2022).

**1.1.2. Autocorrelation**

The autocorrelation function is defined as

$$\rho(\tau) = \frac{\mathbb{E}[\eta(t)\eta(t+\tau)]}{\text{Var}(\eta)}. \tag{13}$$

The numerator is given by

$$\mathbb{E}[\eta(t+\tau)\eta(t)] = \mathbb{E}\left[\left(\sigma e^{\lambda(t+\tau)} \int_0^{t+\tau} e^{-\lambda t''} \xi(t'') dt''\right) \left(\sigma e^{\lambda t} \int_0^t e^{-\lambda t'} \xi(t') dt'\right)\right] \tag{14}$$

$$= \sigma^2 e^{2\lambda t} e^{\lambda \tau} \int_0^{t+\tau} dt'' \int_0^t dt' e^{-\lambda(t''+t')} \delta(t' - t'') \tag{15}$$

$$= \sigma^2 e^{2\lambda t} e^{\lambda \tau} \int_0^{t+\tau} dt'' e^{-2\lambda t''} \tag{16}$$

$$= \frac{\sigma^2 e^{\lambda \tau}}{2\lambda} \left(e^{2\lambda t} - 1\right), \tag{17}$$

which, in the stationary regime ($t \to \infty$) gives an autocorrelation function of

$$\rho(\tau) = e^{\lambda \tau}. \tag{18}$$

Note that as $\lambda \to 0$ from below, the autocorrelation tends to 1, another indication of critical slowing down.

*1.2. Discrete-Time Process*

Some systems are more naturally modeled in discrete time, such as population dynamics with non-overlapping generations. Discrete-time processes also arise as numerical approximations to continuous-time systems when the time step is small.

A general form of a discrete-time stochastic process with additive white noise is

$$x_{t+1} = f(x_t) + \sigma \epsilon_t, \tag{19}$$

where $f$ describes the deterministic dynamics, $\sigma$ is the noise amplitude, and $\epsilon_t \sim \mathcal{N}(0,1)$ are independent and identically distributed (i.i.d.) standard Normal random variables.

Let $x^*$ be a fixed point of the deterministic dynamics, i.e., an equilibrium satisfying $f(x^*) = x^*$. To study fluctuations about equilibrium, define the residual $\eta_t = x_t - x^*$, so that the dynamics are centered around $x^*$. Then

$$\eta_{t+1} = f(x^* + \eta_t) - x^* + \sigma \epsilon_t \tag{20}$$

Taking a first-order Taylor expansion of $f$ about $x^*$ and noting that $f(x^*) = x^*$, we have

$$f(x^* + \eta_t) \approx f(x^*) + f'(x^*)\eta_t = x^* + \alpha \eta_t,$$

where $\alpha = f'(x^*)$. Substituting into the update rule yields

$$\eta_{t+1} = \alpha \eta_t + \sigma \epsilon_t. \tag{21}$$

This is known as an autoregressive process of order 1 (AR(1)). It is the discrete-time analogue of the Ornstein–Uhlenbeck process.

**1.2.1. Variance**

Taking the expectation of both sides of the AR(1) equation,

$$\mathbb{E}[\eta_{t+1}] = \alpha \mathbb{E}[\eta_t] + \sigma \mathbb{E}[\epsilon_t], \tag{22}$$

and using $\mathbb{E}[\epsilon_t] = 0$, we find that the mean satisfies

$$\mathbb{E}[\eta_{t+1}] = \alpha \mathbb{E}[\eta_t].$$

Assuming stationarity, the mean is constant in time, so $\mathbb{E}[\eta_{t+1}] = \mathbb{E}[\eta_t] = \mu$. This gives

$$\mu = \alpha\mu \quad \Rightarrow \quad \mu(1 - \alpha) = 0,$$

implying $\mu = 0$ (for $\alpha \neq 1$). Squaring both sides of the AR(1) equation and taking expectations yields

$$\mathbb{E}[\eta_{t+1}^2] = \alpha^2 \mathbb{E}[\eta_t^2] + 2\alpha\sigma \mathbb{E}[\eta_t \epsilon_t] + \sigma^2 \mathbb{E}[\epsilon_t^2]. \tag{23}$$

Since $\eta_t$ and $\epsilon_t$ are independent and $\mathbb{E}[\epsilon_t] = 0$, we have $\mathbb{E}[\eta_t \epsilon_t] = 0$, and $\mathbb{E}[\epsilon_t^2] = 1$. Thus,

$$\mathbb{E}[\eta_{t+1}^2] = \alpha^2 \mathbb{E}[\eta_t^2] + \sigma^2. \tag{24}$$

Assuming stationarity (i.e., constant variance), we get

$$\mathrm{Var}(\eta) = \alpha^2 \mathrm{Var}(\eta) + \sigma^2, \tag{25}$$

which rearranges to

$$\mathrm{Var}(\eta) = \frac{\sigma^2}{1 - \alpha^2}. \tag{26}$$

This expression is valid only for $|\alpha| < 1$, which ensures that the process is stationary and the variance remains finite. The variance diverges as $\alpha \to 1$ from below, an indication of critical slowing down. This is expression is also given in Scheffer et al. (2009).

**1.2.2. Lag-1 autocorrelation**

Multiplying the AR(1) equation by a factor of $\eta_t$, and taking expectations gives

$$\mathbb{E}[\eta_{t+1}\eta_t] = \alpha \mathbb{E}[\eta_t^2] + \sigma \mathbb{E}[\epsilon_t \eta_t]. \tag{27}$$

Since $\eta_t$ and $\epsilon_t$ are independent and have zero mean, $\mathbb{E}[\epsilon_t \eta_t] = 0$. Thus,

$$\mathbb{E}[\eta_{t+1} \eta_t] = \alpha \mathbb{E}[\eta_t^2]. \tag{28}$$

Dividing both sides by $\mathbb{E}[\eta_t^2]$, we obtain the lag-1 autocorrelation:

$$\rho(1) = \frac{\mathbb{E}[\eta_{t+1} \eta_t]}{\mathbb{E}[\eta_t^2]} = \alpha. \tag{29}$$

The lag-1 autocorrelation tends to one as $\alpha \to 1$ from below, another indication of critical slowing down. Note that here, the AR(1) coefficient ($\alpha$) is equal to the lag-1 autocorrelation. This is not the case in general (as a stochastic process may not be well described by an AR(1) process in general). However, systems at equilibrium subject to small perturbations are well described by an AR(1) process, hence the AR(1) coefficient can also serve as an early warning signal for systems at equilibrium, gradually approaching a bifurcation.